\documentclass[10pt,sn-basic]{elsarticle}
\usepackage{flafter}
\usepackage{graphicx}
\usepackage{subcaption}
\usepackage{amsmath}
\usepackage{xcolor}
\usepackage{array}
\usepackage{amssymb}
\usepackage{amsmath}
\usepackage{graphicx}
\usepackage{ulem}
\newcolumntype{P}[1]{>{\centering\arraybackslash}p{#1}}
\makeatletter
\newcommand*{\centerfloat}{%
  \parindent \z@
  \leftskip \z@ \@plus 1fil \@minus \textwidth
  \rightskip\leftskip
  \parfillskip \z@skip}
\makeatother
\graphicspath{ {./images/} }
\begin{document}
\begin{frontmatter}
\title{First observation of the cosmic ray shadow of the Moon and the Sun with KM3NeT/ORCA}
\cortext[cor]{corresponding author}

\author[a]{S.~Aiello}
\author[bd,b]{A.~Albert}
\author[c]{S. Alves Garre}
\author[d]{Z.~Aly}
\author[e,f]{A. Ambrosone}
\author[g]{F.~Ameli}
\author[h]{M.~Andre}
\author[i]{M.~Anghinolfi}
\author[j]{M.~Anguita}
\author[k]{M. Ardid}
% ...
\address[a]{INFN, Sezione di Catania, Via Santa Sofia 64, Catania, 95123 Italy}
\address[b]{Universit{\'e}~de~Strasbourg,~CNRS,~IPHC~UMR~7178,~F-67000~Strasbourg,~France}
\address[c]{IFIC - Instituto de F{\'\i}sica Corpuscular (CSIC - Universitat de Val{\`e}ncia), c/Catedr{\'a}tico Jos{\'e} Beltr{\'a}n, 2, 46980 Paterna, Valencia, Spain}
\address[d]{Aix~Marseille~Univ,~CNRS/IN2P3,~CPPM,~Marseille,~France}
\address[e]{INFN, Sezione di Napoli, Complesso Universitario di Monte S. Angelo, Via Cintia ed. G, Napoli, 80126 Italy}
\address[f]{Universit{\`a} di Napoli ``Federico II'', Dip. Scienze Fisiche ``E. Pancini'', Complesso Universitario di Monte S. Angelo, Via Cintia ed. G, Napoli, 80126 Italy}
\address[g]{INFN, Sezione di Roma, Piazzale Aldo Moro 2, Roma, 00185 Italy}
\address[h]{Universitat Polit{\`e}cnica de Catalunya, Laboratori d'Aplicacions Bioac{\'u}stiques, Centre Tecnol{\`o}gic de Vilanova i la Geltr{\'u}, Avda. Rambla Exposici{\'o}, s/n, Vilanova i la Geltr{\'u}, 08800 Spain}
\address[i]{INFN, Sezione di Genova, Via Dodecaneso 33, Genova, 16146 Italy}
\address[j]{University of Granada, Dept.~of Computer Architecture and Technology/CITIC, 18071 Granada, Spain}
\address[k]{Universitat Polit{\`e}cnica de Val{\`e}ncia, Instituto de Investigaci{\'o}n para la Gesti{\'o}n Integrada de las Zonas Costeras, C/ Paranimf, 1, Gandia, 46730 Spain}
% ...
\address[bd]{Universit{\'e} de Haute Alsace, rue des Fr{\`e}res Lumi{\`e}re, 68093 Mulhouse Cedex, France}
% ...

\vspace{30mm} 
\begin{abstract}
This article reports the first observation of the Moon and the Sun shadows in the sky distribution of cosmic-ray induced muons measured by the KM3NeT/ORCA detector. The analysed data-taking period spans from February 2020 to November 2021, when the detector had 6 Detection Units deployed at the bottom of the Mediterranean Sea, each composed of 18 Digital Optical Modules. The shadows induced by the Moon and the Sun were detected at their nominal position with a statistical significance of 4.2$\sigma$ and 6.2$\sigma$,  and an angular resolution of $\sigma_{res}=0.49^\circ$ and $\sigma_{res}=0.66^\circ$, respectively, consistent with the prediction of $0.53^\circ$ from simulations. This early result confirms the effectiveness of the detector calibration, in time, position and orientation and the accuracy of the event direction reconstruction. This also demonstrates the performance and the competitiveness of the detector in terms of pointing accuracy and angular resolution.
\end{abstract}

\end{frontmatter}
\tableofcontents
\newpage
\section{Introduction}

\begin{figure}[!htbp]
 \includegraphics[scale=0.5]{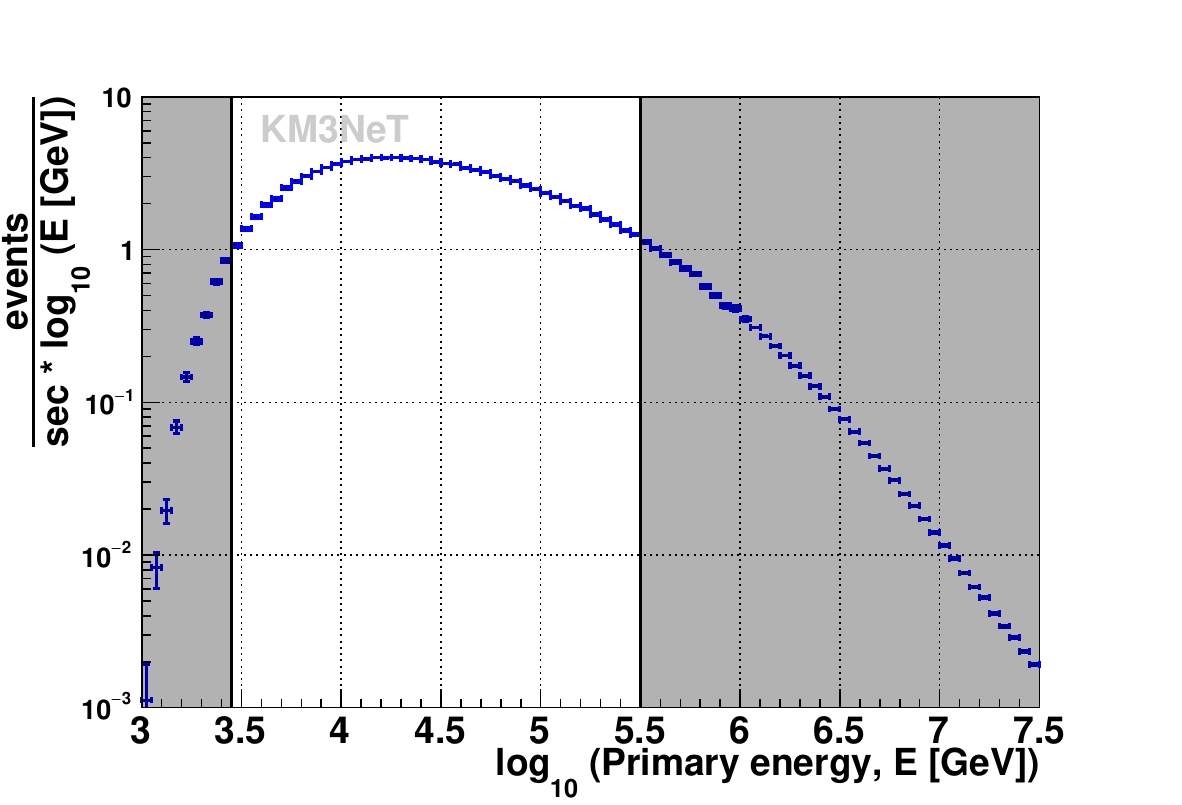}
  \centerfloat 
 \caption{Distribution of CR events which yield a reconstructed event in the KM3NeT/ORCA detector. The white region indicates the energy range which provides 90\% of the event sample.}
 \label{Energy_CR}
\end{figure}

Cosmic rays (CR) are charged particles mainly composed of protons and light nuclei. Since they can be deflected by irregular Galactic magnetic fields, their arrival directions at the Earth are almost isotropic. Once a primary CR particle reaches the Earth's upper atmosphere and interacts with an air nucleus, it produces secondary particles. The most penetrating component of these are muons, which can be detected at the surface of the Earth but also at underground or underwater detectors.  In order to reach the KM3NeT/ORCA detector~\cite{LOI} at a depth of 2500 meters below sea level~\cite{muons}, vertically down-going muons need a minimal energy of around 500 GeV at sea level and thus they must have originated from primary CRs of energies exceeding several TeV/nucleon as described in \cite{runbyrun}.  90\% of the CR particles yielding muons used in this analysis have energies between 3 and 330 TeV as illustrated in Figure~\ref{Energy_CR}. The angle between the primary CR and the secondary muon is on average within 0.1° at these energies~\cite{angle} which is around 5 times lower than the expected angular resolution of the detector for muons. At the KM3NeT/ORCA detector the muons are dominantly minimal-ionizing with energies ranging from few tens of GeV to few hundred GeV.
\newline CRs are blocked by nearby celestial bodies such as the Moon and the Sun. This induces a deficit in the atmospheric muon flux and in other secondary CR particles coming from the direction of these objects. This effect had been predicted by Clark in 1957~\cite{clark}. Its observation can be used to verify the pointing accuracy and angular resolution of detectors which are able to measure secondary particles from CR interactions. The Moon and the Sun shadows in CRs have been observed by multiple experiments (IceCube \cite{IC_sun}, ANTARES \cite{antares_moon, antares_sun},  MACRO \cite{ambrosio}, L3~\cite{achard}, MINOS~\cite{adamson}, ARGO-YBJ~\cite{bartoli}, HAWC~\cite{abeysekara} and others). As a permanent bright high-energy neutrino source has yet to be found, the observation of the Moon and Sun shadows in CRs is an important calibration tool for neutrino telescopes such as KM3NeT/ORCA and helps to demonstrate their pointing accuracy and to measure their angular resolution.

\section{KM3NeT/ORCA detector}
KM3NeT is a research infrastructure consisting of undersea Cherenkov neutrino telescopes currently under construction at the bottom of the Mediterranean Sea off-shore the Italian Sicily coast (KM3NeT/ARCA) and 40 km off-shore Toulon, France (KM3NeT/ORCA)~\cite{LOI}. The two detectors are optimised for different neutrino energy ranges. They are composed of vertical Detection Units (DUs). Six of them had been operational in KM3NeT/ORCA when the data used in this analysis were acquired. Each DU consists of 18 spherical Digital Optical Modules (DOMs), with 31 photomultiplier tubes (PMTs) distributed almost isotropically within each DOM \cite{dom}. These PMTs detect the Cherenkov light emitted along the path of relativistic charged particles propagating through water. In the data acquisition a \textit{hit} is produced when a photon impinging on a PMT induces an electrical signal above a defined threshold. A hit consists of a time stamp and a time over threshold. An event is created when the trigger algorithm identifies a series of causally-connected hits. For the analysis presented here, these events are processed by a track reconstruction algorithm. The hits are fitted with a model of a Cherenkov light emitting muon. This particle is assumed to follow a long, straight trajectory and to propagate practically at the speed of light in vacuum through water. The position, time and direction of such a track is determined by using a maximum-likelihood method based on a set of causally-connected hit times and positions~\cite{reco}. A time-dependent calibration of the detector, with the monitored positions and orientations of every DOM computed and interpolated every 10 minutes, is used in this work to account for movements of the strings with the sea current. 

\section{Data and Monte Carlo samples}

\begin{figure}[!htbp]
 \includegraphics[scale=0.2]{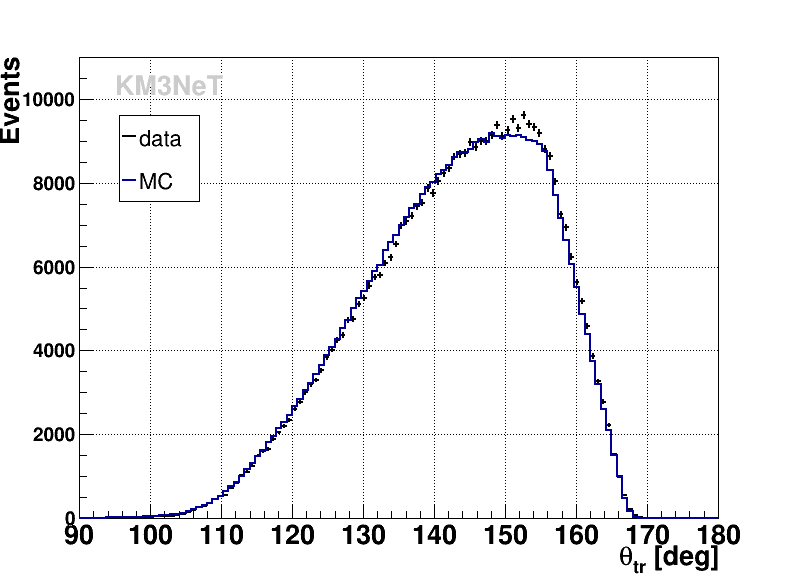} 
 \includegraphics[scale=0.2]{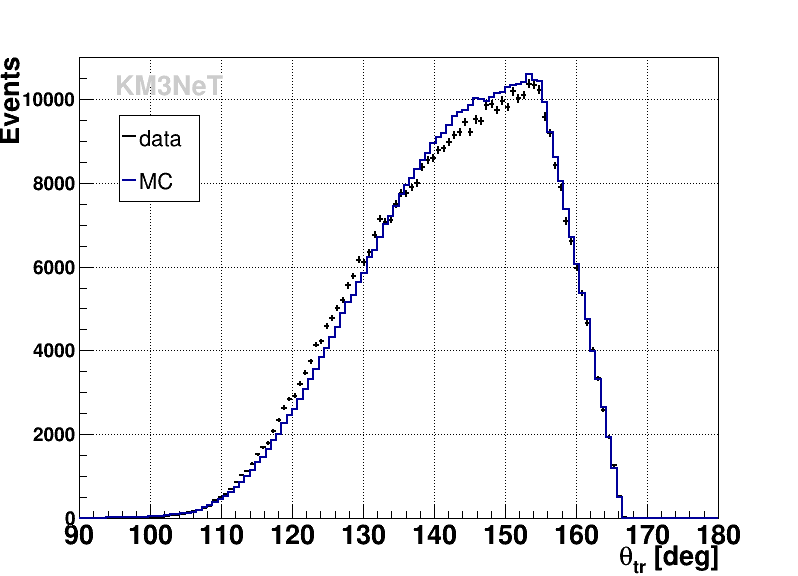}
 \centerfloat 
 \caption{Zenith angle distribution of selected events for the Moon (left) and the Sun (right). Data (black crosses) and simulations rescaled to data (blue histograms) are shown.}
 \label{zenith}
\end{figure}

The data used in this analysis were collected between February 11, 2020, and November 18, 2021 for a total of 499.3 days. Quality cuts on  the number of used hits and the likelihood of the track reconstruction were applied to remove poorly reconstructed events keeping 83\% of the initial event sample. An average event density of 3000 events per square degree is measured  in the vicinity of Moon and Sun. It is expected that about 640 CR events are blocked by each of the two sky objects.   The position of the Moon/Sun in the sky is obtained using the \textit{astropy} package \cite{astropy} that relies on the International Celestial Reference System (ICRS) coordinates described in \cite{ICRS}. 
The latitude of $43^\circ$ North of the detector and the data taking period of more than one year lead to broad zenith angle distributions of the selected events in the vicinity of Sun and Moon as shown in Figure~\ref{zenith}.

Monte Carlo (MC) simulations are used to optimise 
the track selection, 
and predict the angular resolution and amplitude of the CR deficit induced by the Moon and the Sun. Secondary muons at the KM3NeT/ORCA detector are simulated with the MUPAGE package~\cite{mupage}~\cite{mupage2}. These muons are then propagated through sea water and Cherenkov photons are created within a cylindrical volume surrounding the simulated detector. Finally the detector response is simulated by producing digitized hits from photons detected by PMTs and by adding noise hits from environmental background mainly due to $^{40}$K decay and bioluminescence. These are derived  from real data runs in a time-dependent way following the \textit{run-by-run} approach previously introduced by the ANTARES Collaboration \cite{runbyrun}. The time-dependent PMT efficiencies are monitored and used in the simulation. The last two steps are done with KM3NeT custom software~\cite{reco}. The resulting hit patterns are passed through the trigger and reconstruction steps using the same software that is used to process real data. The simulated track sample is about 4 times larger compared to the real data sample.

The optimisation of the track selection is performed by varying the cut on the angular error estimate from the track reconstruction to maximize the significance of the shadow signal.
By requiring a more and more stringent cut  a smaller but higher quality event sample is selected.
The chosen value keeps 45\% of the original MC sample.

\label{sec}

\section{Analysis method}
The search for the shadow of the Moon/Sun is done in the phase space of the angular differences between the reconstructed track coordinates and the celestial object. Both a 1-dimensional and a 2-dimensional analysis have been performed. The 1-dimensional analysis uses the space angle between the direction of the Moon/Sun and the track.  The 2-dimensional analysis uses two Cartesian coordinates $(x,y)$, starting from the zenith and azimuth angles of the sky object $\theta_{sk}, \alpha_{sk}$ and the tracks $\theta_{tr}, \alpha_{tr}$ respectively,
\begin{equation}
\begin{split}
    &x =  (\alpha_{sk} - \alpha_{tr})\sin\theta_{tr} \\
    &y =  \theta_{sk}-\theta_{tr} \\
\end{split}
\end{equation}
2D maps that extend to $\pm 6^\circ$ in $(x,y)$ have been constructed. Data are only used for times when the full $\pm 6^\circ$ field around the Moon/Sun is above the horizon. An angular range of $\pm 6^\circ$ contains the detectable shadow signal entirely and it allows for a simultaneous fit of the shadow and background parameters while analysing just a moderately-sized atmospheric muon sample. It has been verified that the results do not depend on the precise choice of the chosen angular range. A constant $0.1^\circ$ binning in $x$ and $y$ is chosen, however results are reproduced when choosing a smaller binning. The 1D histogram contains tracks that are closer than $4^\circ$ to the Moon/Sun position.

The significance of the shadowing effect of Moon/Sun is determined with a likelihood ratio test, by comparing the likelihood of a background hypothesis model $H_0$, with the likelihood of a signal plus background model $H_1$ which includes a shadowing effect.
The Poisson likelihood with the definition in Ref~\cite{pois} \\
\begin{equation}
\chi^2(H) = 2\sum_{i}^{N_{bin}}[N_{i,H}-n_i+n_i \mbox{ln}(n_i/N_{i,H})]
 \label{chi}
\end{equation}
is used, where $n_i$ stands for the event count in the $i$-th space angle bin to be compared with the expectations $N_{i,H}$ under the $H_0$ and $H_1$ hypotheses. The difference in $\Delta\chi_{H1/H0}^2 = \chi^2(H_1)-\chi^2(H_0)$ values is used to determine the probability to reject the null hypothesis and to extract the significance of the observation from it.  The background event distribution in azimuth is found to be uniform, while the zenith angle dependency can be conveniently parametrized with a $2^{nd}$ order polynomial function. This yields for the 2D maps
\begin{equation}
N_{i,H_0} =  \rho \big[1+a_1 y_i + a_2 y_i^2\big] 
\label{H0}
\end{equation}
 with $\rho$ a constant track density per space angle and $a_1, a_2$ parameters which are determined during minimisation. For the 1D maps, each bin contains events from an almost symmetric zenith angle range above and below the Moon/Sun position resulting in a uniform exposure, {\it i.e.} $N_{i,H_0} = \rho$.

The event expectation in bin $i$ for the signal plus background hypothesis $H_1$ is defined as
\begin{equation}
    N_{i,H_1} = N_{i,H_0} - \rho \cdot G_i
\end{equation}
where $G_i$ describes the deficit of secondary CR events due to the shadowing effects of Moon/Sun in bin $i$ with coordinates $x_i$, $y_i$ as a bi-dimensional Gaussian
\begin{equation}
  G_i(A,\sigma_{res}, x_s, y_s) =  A\frac{R_s^2}{2\sigma_{res}^2}\exp\Big[-\frac{(x_i-x_s)^2+(y_i-y_s)^2}{2\sigma_{res}^2}\Big] 
 \label{H1}
\end{equation}
where $A$ is the relative shadow amplitude. For $A=1$ the number of blocked CR events correspond to $\rho \pi R_s^2$. The angular resolution of the detector for the selected sample of CR events is measured by $\sigma_{res}$, the angular width of the Gaussian shadow. The $R_s$, $x_s$, $y_s$ parameters are the apparent angular radius of the celestial object and the relative angular position of the Moon/Sun shadow with respect to their nominal positions. For the 1D maps, the term 
$(x_i-x_s)^2 + (y_i-y_s)^2$ in Equation \ref{H1} is replaced by $\delta^2$, the square of the angular distance between the track and the sky object from Equation~\ref{space angle} with $V_{tr}$ and $V_{sk}$ the direction vector of the track and the sky object respectively.
\begin{equation}
\delta = \arccos{ ( V_{tr} \cdot V_{sk} ) } * 180 / \pi
\label{space angle}
\end{equation}
The significance of the shadow is found by fitting $\rho$, $a_1$,  $a_2$, $A$ and $\sigma_{res}$ at $(x_s,y_s)=(0,0)$ on the 2D or 1D event distribution respectively. Results from these fits are summarized in section 5.2. The position of the shadow is obtained by simultaneously fitting $\rho$, $a_1$,  $a_2$, $(x_s,y_s)$ and $A$, with $\sigma_{res}$ fixed to its expectation value from MC. Results from these fits are found in Section 5.3.

\begin{figure}[!htbp]
 \includegraphics[scale=0.278]{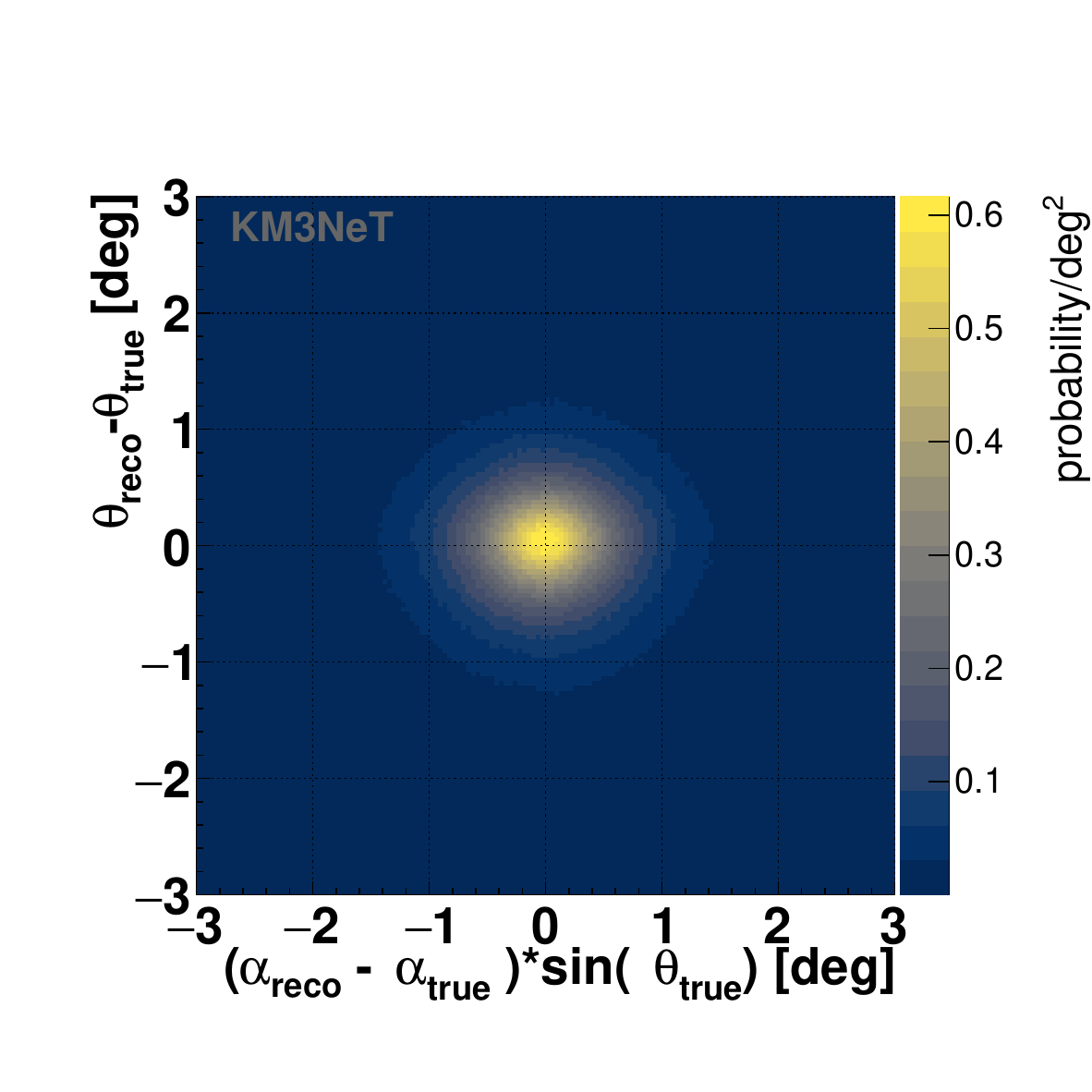} 
 \includegraphics[scale=0.408]{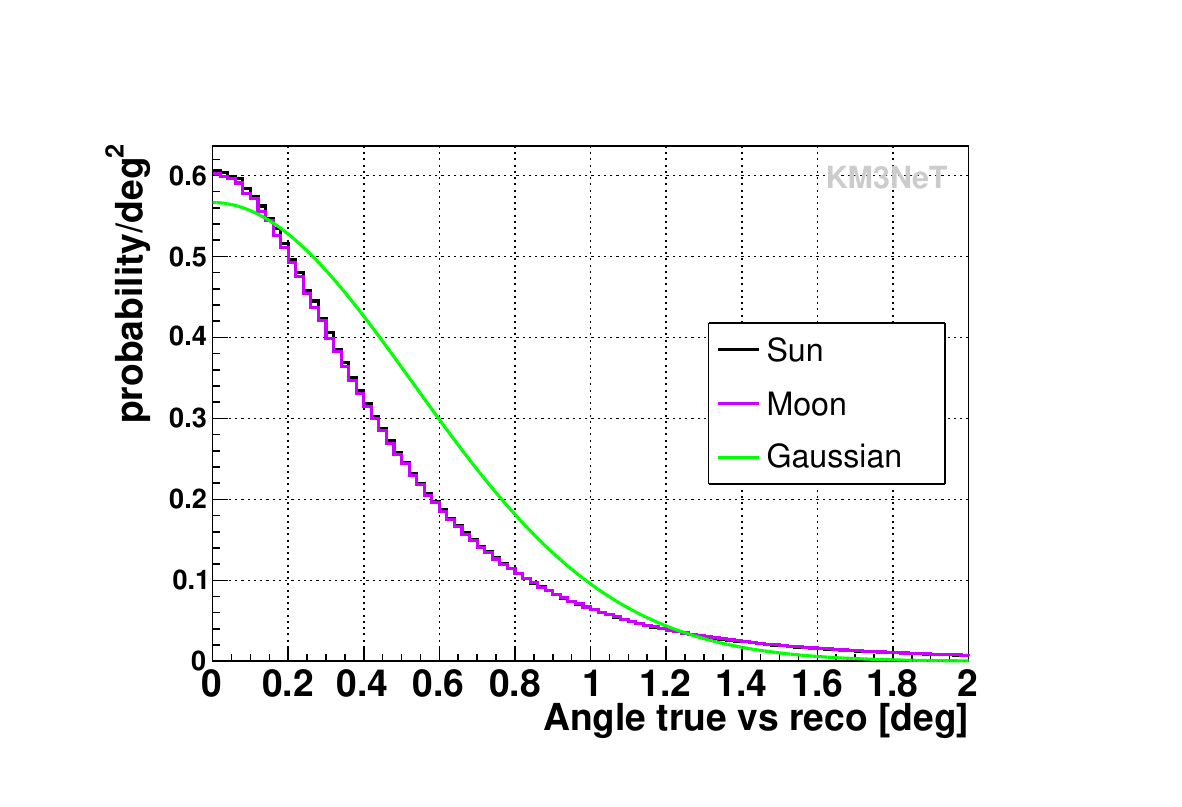}
 \centerfloat 
 \caption{PSF from MC, {\it i.e.} the difference between the true muon direction and the reconstructed one in 2-dimensional angular space (left) and as function of the space angle difference (right, blue and magenta for Sun and Moon, respectively). A Gaussian PSF with $\sigma=0.53^\circ$ is shown in green for comparison.}
 \label{psf}
\end{figure}

The assumption of Gaussianity of the shadowing effect is the result of a few approximations. Firstly, the influence of the size of the Moon/Sun is neglected. This is acceptable as long as the angular resolution of the detector is larger than the angular radius of the Moon/Sun, a condition which is amply satisfied in the present case. In addition, the real point spread function (PSF) of the detector 
 has a slightly different radial shape compared to a Gaussian function (see Figure~\ref{psf}, right) which
will be accounted for by fitting the shadow amplitude $A$ when comparing data to the $H_1$ hypothesis. It is observed, that the PSF is identical for the data samples selected for Moon and Sun. 

Further, the PSF is perfectly symmetric in $x$ and $y$ as shown in Figure~\ref{psf} (left), allowing for a reliable fit of $x_s$ and $y_s$.

The Moon radius varies between $0.245^\circ$ and $0.279^\circ$ resulting in an amplitude variation of $\pm 14\%$ around the mean value. However our data sample covers several Moon cycles,
so the average value is used. A similar statement can be made for the Sun whose apparent radius varies between $0.262^\circ$ and $0.271^\circ$ during the year.

\section{Data analyses}
\subsection{Background}
\begin{figure}[!htbp]
 \includegraphics[scale=0.27]{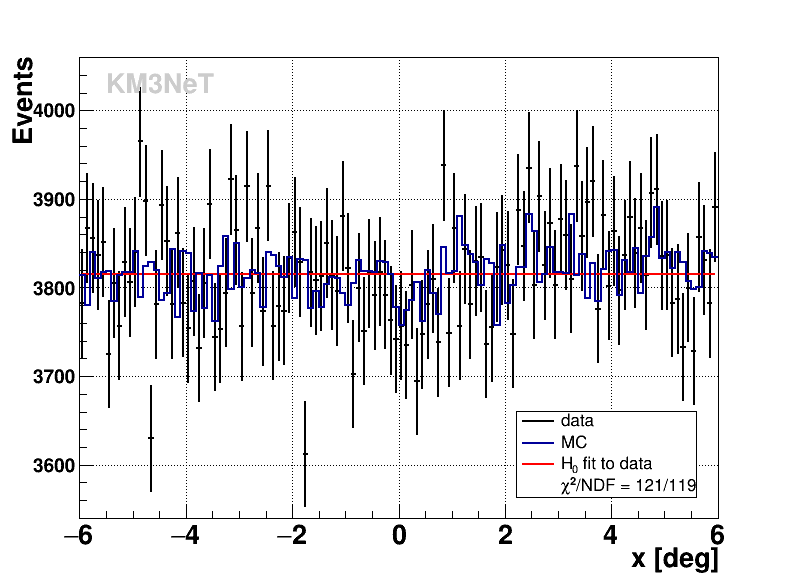} 
 \includegraphics[scale=0.27]{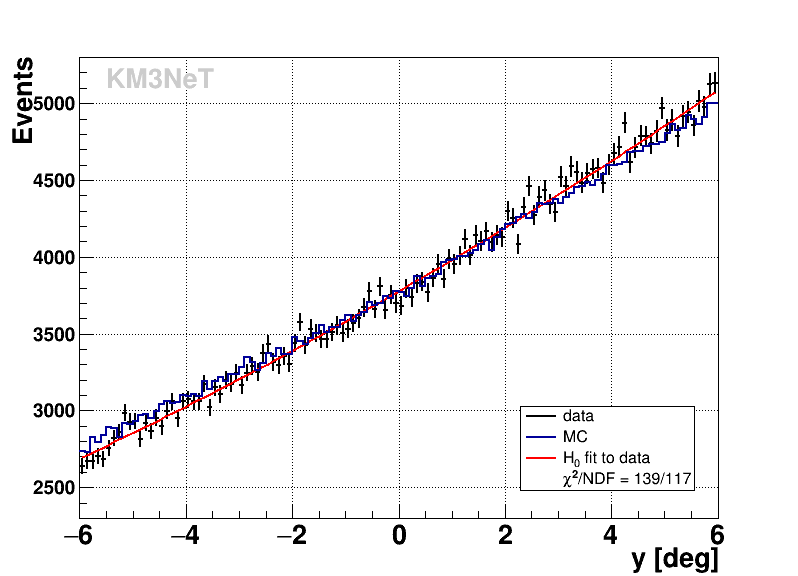}
 \centerfloat 
 \caption{Event distribution in $x$ and $y$ for the Sun data sample (black crosses) compared to the polynomial fits from the $H_0$ hypothesis (red lines)
  and MC predictions rescaled to data (blue histograms)}.
 \label{data_sun_az_zen}
\end{figure}

The background distributions in $x$ and $y$ are shown in
Figure~\ref{data_sun_az_zen} for the Sun as an example. Fits of the $H_0$ hypothesis using a constant, and $2^{nd}$ order polynomial function describe the data well. The MC predictions are also compatible with these functions.

\newpage

\subsection{Fits at nominal positions}

\begin{table}[htp]
\begin{center}
\centerfloat
    \begin{tabular}{|c|c|c|c|c|c|}
    \hline
    Parameters & Moon 1D &  Moon 2D & Sun 1D & Sun 2D\\
    \hline
    $\sigma_{res}$ & $0.49^\circ \pm 0.11^\circ$ & $0.49^\circ \pm 0.15^\circ$ & $0.66^\circ \pm 0.08^\circ$ & $0.65^\circ\pm0.13^\circ$\\
    $A$ & $0.69 \pm 0.17$ & $0.71 \pm 0.27$ & $1.38 \pm 0.31$ & $1.31 \pm 0.34$\\
    $\Delta\chi_{H1/H0}^2$ & -20.7  & -21.3 & -47.2 & -43.0\\
    Significance   & $4.2\sigma$ & $4.2\sigma$ & $6.5\sigma$ & $ 6.2\sigma$ \\
    Events/deg$^2$ & $2886$ & $2892$ & $3166$ & $3161$ \\
    \hline
    \end{tabular}
    \caption{Parameters from the fits at nominal position $(x_s,y_s)=(0,0)$ }
    \label{par}
\end{center}
\end{table}

\begin{figure}[!htbp]
    \centerfloat 
    \includegraphics[scale=0.27]{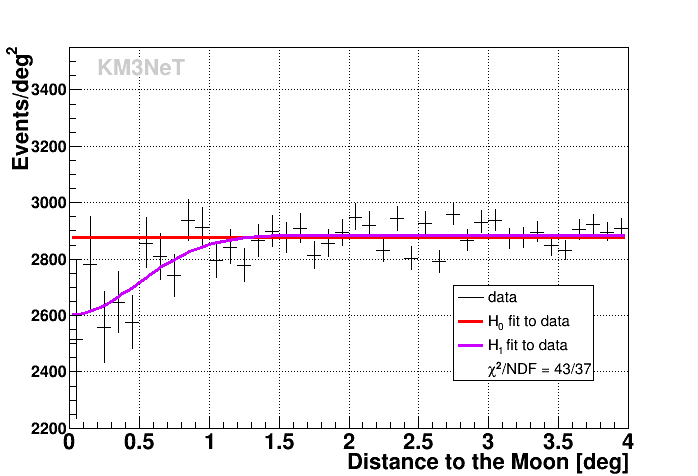}
    \includegraphics[scale=0.27]{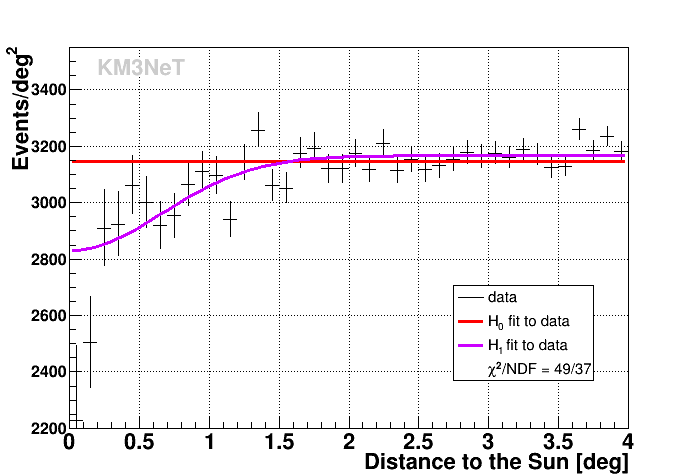}
    \caption{Event density as a function of the distance to the Moon on the left and the Sun on the right. Data (black crosses) are compared to the $H_0$ fit (red) and the $H_1$ fit (magenta).}
    \label{1D}
\end{figure}

Figure~\ref{1D} shows the 1D distributions of the event density as a function of the angular distance from the Moon/Sun. The fit results from the 1D and 2D fits at $(x_s,y_s)=(0,0)$ are summarized in Table ~\ref{par}. 
The significances are derived from the $\Delta\chi_{H1/H0}^2$ with two degrees of freedom ($A$, $\sigma_{res}$). 
 The values obtained for $A$ and $\sigma_{res}$ can be compared to MC expectations of $A=0.90 \pm 0.09$ and $\sigma_{res} = (0.53 \pm 0.04)^\circ$ for the combined Moon/Sun sample.
The fitted values of $\sigma_{res}$ are found compatible with the prediction from simulations. The deeper amplitude and higher significance of the Sun shadow is consistent with the effects of the particular structure of the Sun’s magnetic field during the periods of low solar activity, whose dipole shape is expected to enhance the Sun shadowing effect \cite{sun_variation}.\\

\subsection{Positional fits}

\begin{figure}[!htbp]
        \includegraphics[scale=0.40]{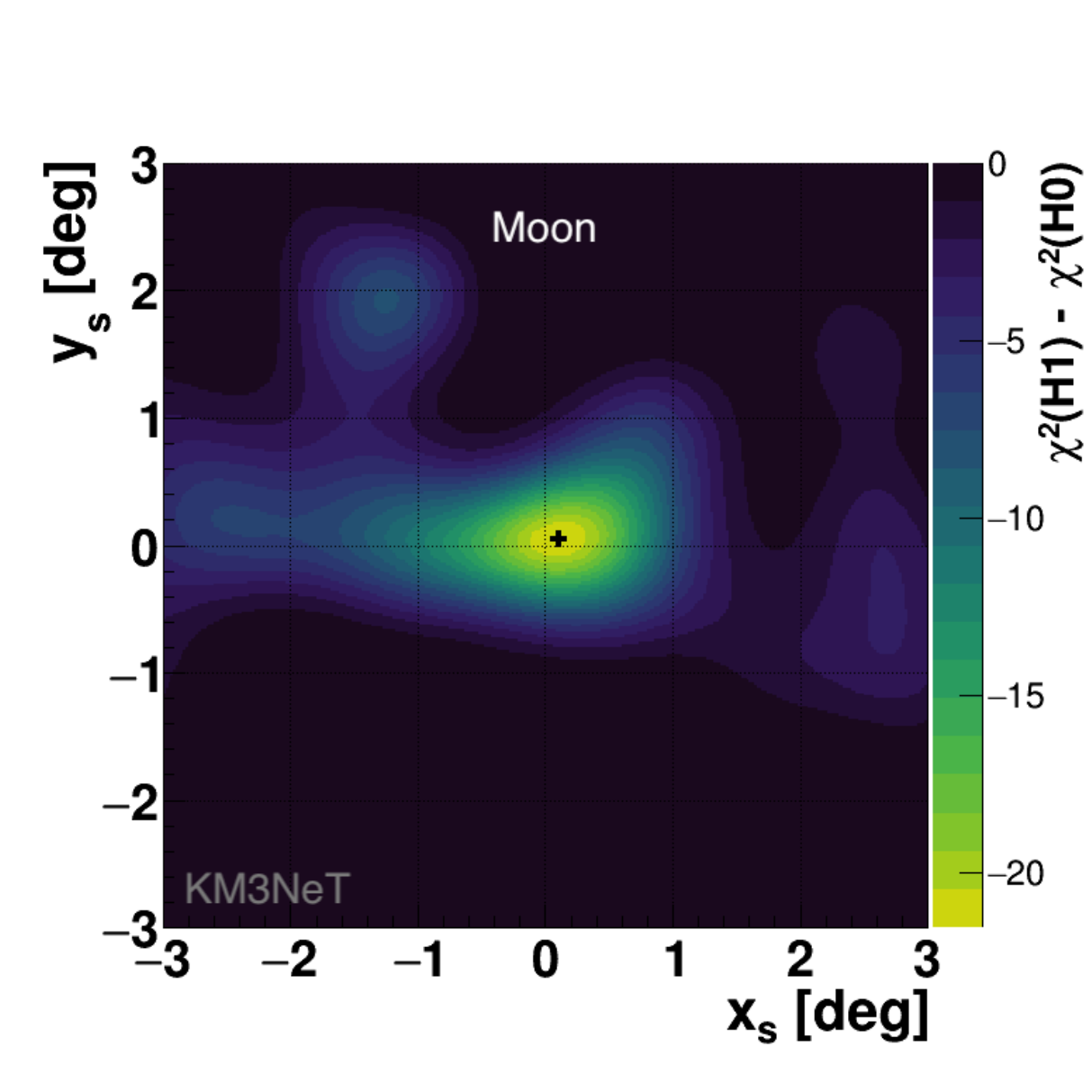}
        \includegraphics[scale=0.40]{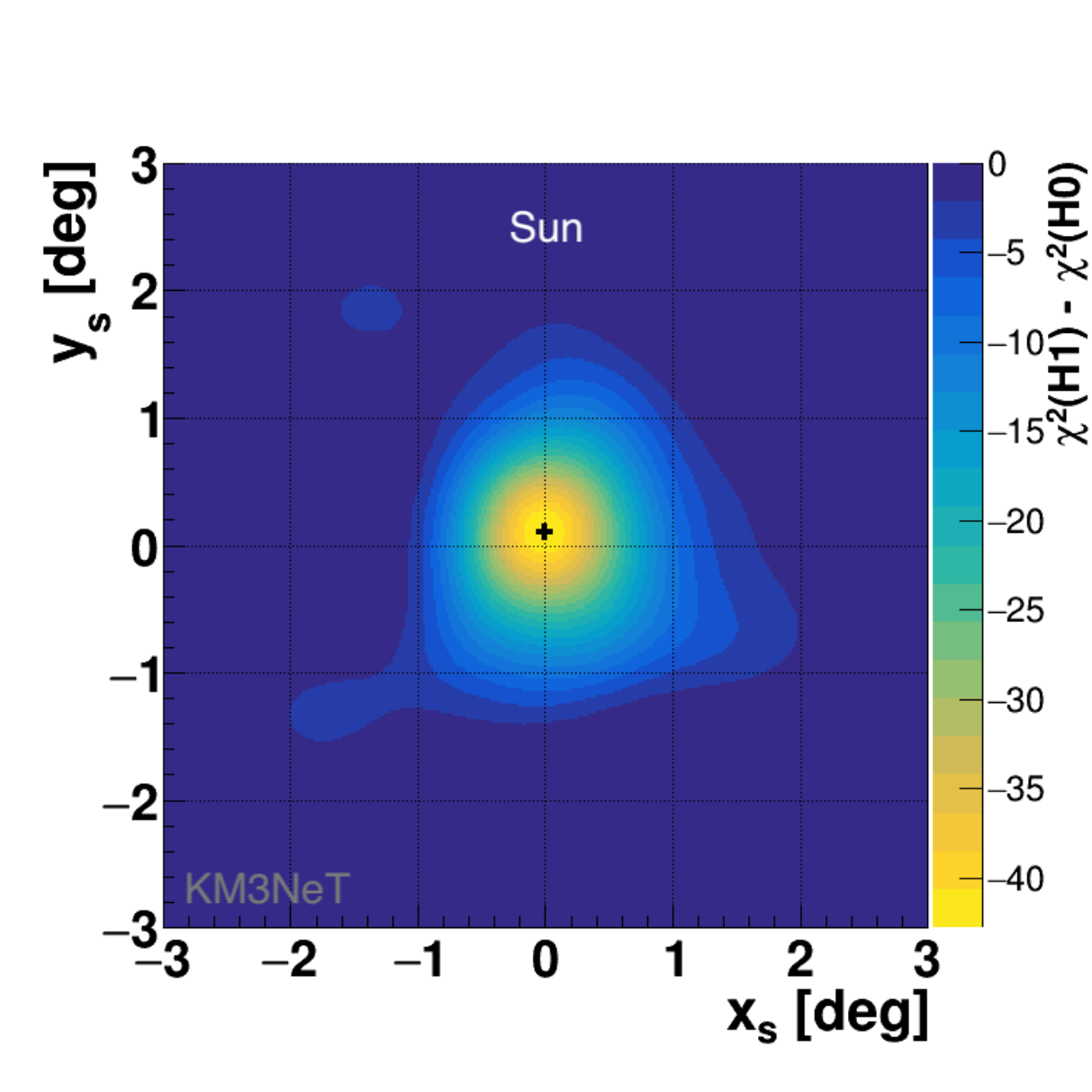}
        \centerfloat
        \caption{colour-coded $\Delta\chi_{H1/H0}^2$ as a function of $x_s$ and $y_s$ for the Moon (left) and the Sun (right).}
        \label{2D}
\end{figure}

\begin{figure}[!htbp]
        \includegraphics[trim={0 0 0 1cm},clip,scale=0.40]{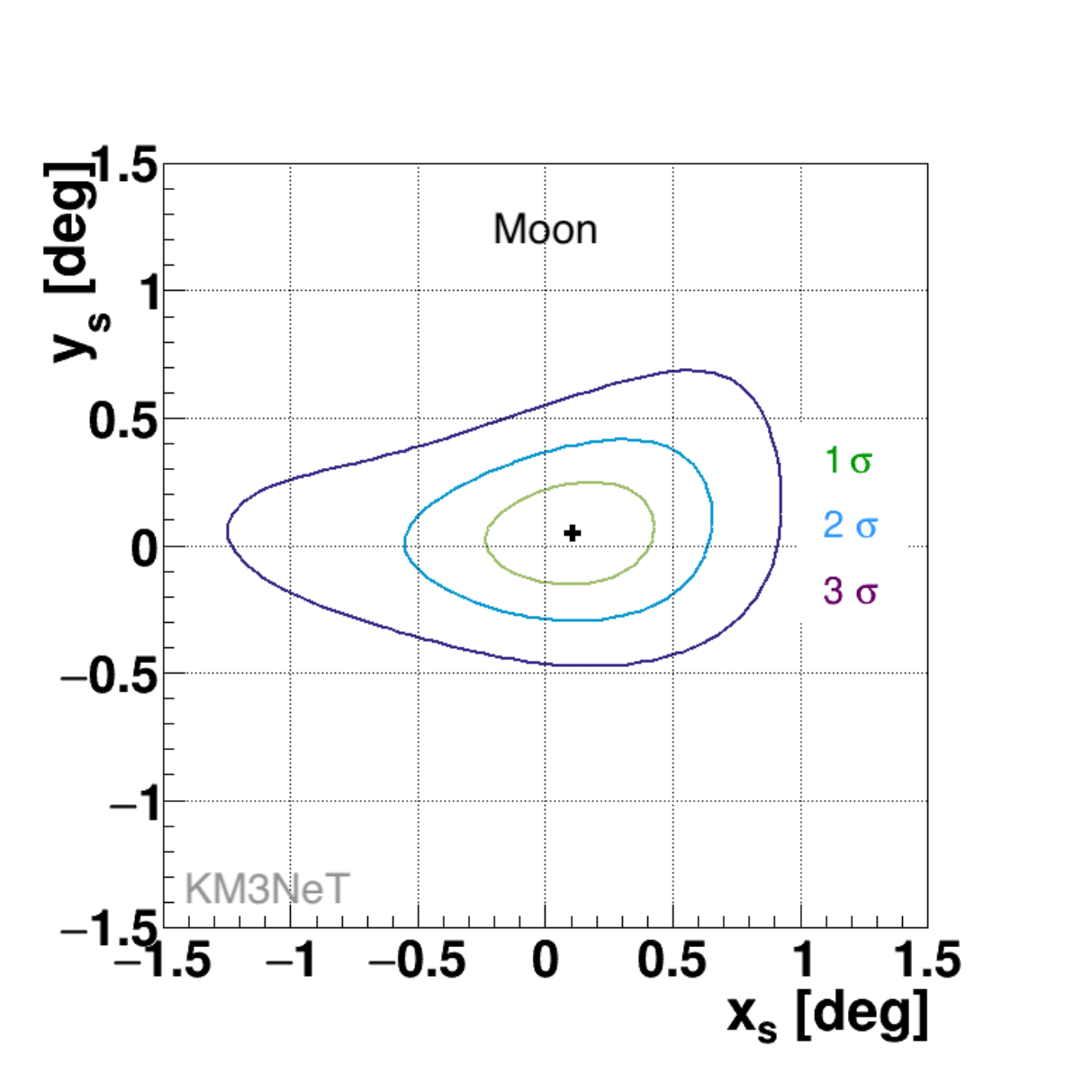}
         \includegraphics[trim={0 0 0 1cm},clip,scale=0.40]{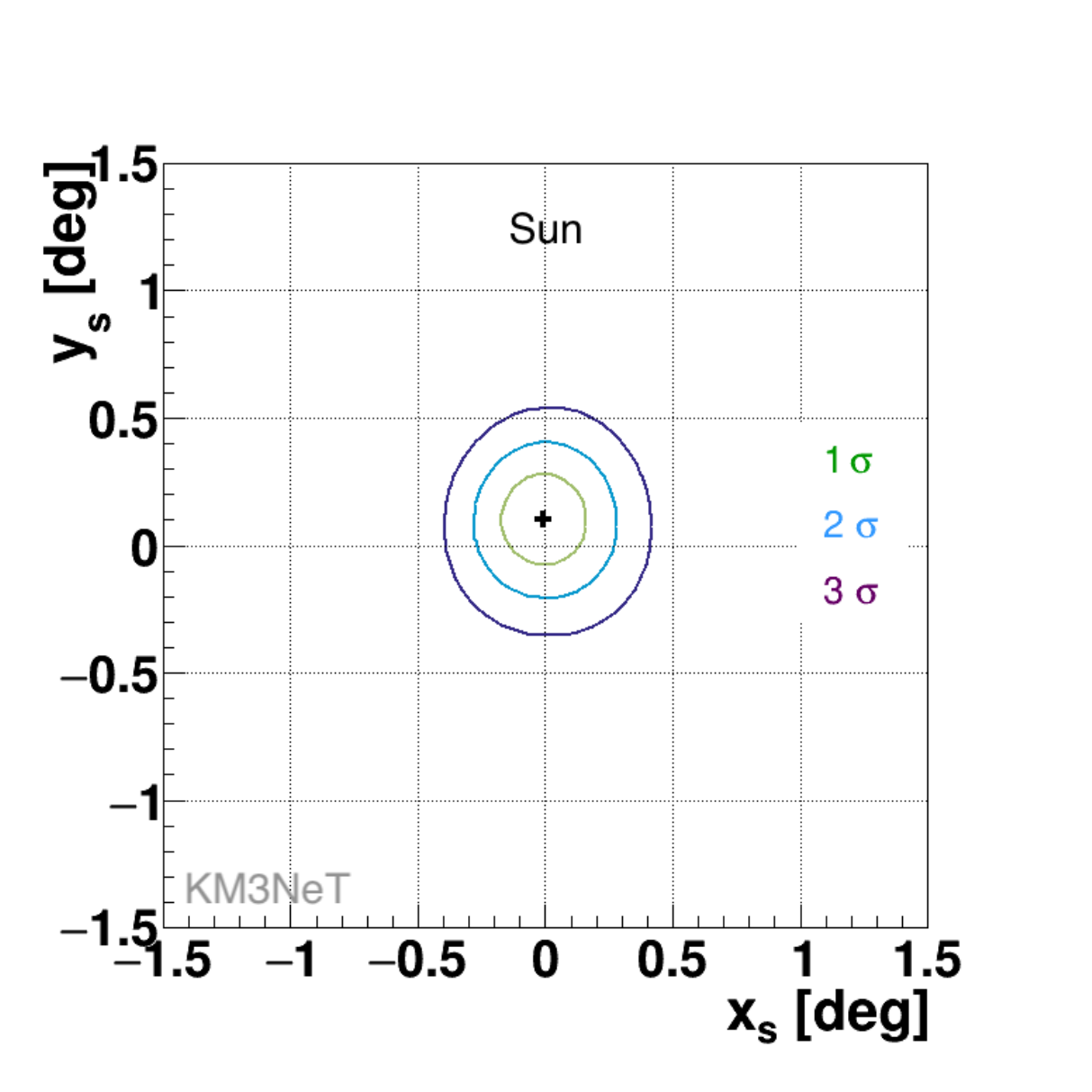}
        \centerfloat
        \caption{Confidence intervals derived from ~Figure \ref{2D} for the Moon (left) and the Sun (right). The black cross indicates the best fit point. }
        \label{cont}
\end{figure}

The results from the fits of the 2D map in $(x_s,y_s)$ with $\sigma_{res}=0.53^\circ$ are shown in Figure~\ref{2D}. The plot illustrates $\chi^2(H_1)-\chi^2(H_0)$ in colour coding. The image of the shadow is clearly visible. Figure~\ref{cont} shows the 1$\sigma$ ($68.3\%$), 2$\sigma$ ($95.4\%$), 3$\sigma$ ($99.7\%$) confidence contours for the two parameters $(x_s,y_s)$ around the best fit point $x_s = (0.11 \pm 0.21)^\circ$, $y_s = (0.04 \pm 0.13)^\circ$ for the Moon and $x_s = (-0.01 \pm 0.11)^\circ$, $y_s = (0.10 \pm 0.12)^\circ$ for the Sun. The true position of the Moon and the Sun in Figure~\ref{cont} are contained within the 68$\%$ contours, yielding a $84\%$ and $67\%$ compatibility between the nominal and the best fit positions, calculated from the corresponding $\Delta\chi^2$ with two degrees of freedom ($x_s$, $y_s$). The slightly different shapes of the contours for Moon and Sun can be entirely attributed to statistical fluctuations.

\section{Conclusion}

The Moon and the Sun's CR shadows have been observed with a high statistical significance using data collected between February 2020  and November 2021 with 6 Detection Units of the KM3NeT/ORCA detector. 
The demonstrated sensitivity to the shadow observation with only one and half years of data taking and a yet incomplete detector reflects the good understanding of the detector positioning, orientation, time calibration \cite{cal} and reconstruction capabilities. The shadow observed in data is compatible with expectations from MC concerning the significance, angular width and amplitude, except for the Sun's amplitude, where it was found above expectations.
\\
\section*{Acknowledgements} The authors acknowledge the financial support of the funding agencies:


Agence Nationale de la Recherche (contract ANR-15-CE31-0020), Centre National de la Recherche Scientifique (CNRS), Commission Europ\'eenne (FEDER fund and Marie Curie Program), Institut Universitaire de France (IUF), LabEx UnivEarthS (ANR-10-LABX-0023 and ANR-18-IDEX-0001), Paris \^Ile-de-France Region, France;
Deutsche Forschungsgemeinschaft (DFG), Germany;
The General Secretariat of Research and Technology (GSRT), Greece;
Istituto Nazionale di Fisica Nucleare (INFN), Ministero dell'Universit\`a e della Ricerca (MIUR), PRIN 2017 program (Grant NAT-NET 2017W4HA7S) Italy;
Ministry of Higher Education, Scientific Research and Innovation, Morocco, and the Arab Fund for Economic and Social Development, Kuwait;
Nederlandse organisatie voor Wetenschappelijk Onderzoek (NWO), the Netherlands;
The National Science Centre, Poland (2015/18/E/ST2/00758);
National Authority for Scientific Research (ANCS), Romania;
Grants PID2021-124591NB-C41, -C42, -C43 funded by MCIN/AEI/ 10.13039/501100011033 and, as appropriate, by “ERDF A way of making Europe”, by the “European Union” or by the “European Union NextGenerationEU/PRTR”,  Programa de Planes Complementarios I+D+I (refs. ASFAE/2022/023, ASFAE/2022/014), Programa Prometeo (PROMETEO/2020/019) and GenT (refs. CIDEGENT/2018/034, /2019/043, /2020/049. /2021/23) of the Generalitat Valenciana, Junta de Andaluc\'{i}a (ref. SOMM17/6104/UGR, P18-FR-5057), EU: MSC program (ref. 101025085), Programa Mar\'{i}a Zambrano (Spanish Ministry of Universities, funded by the European Union, NextGenerationEU), Spain;

%
\bibliographystyle{spphys}
\bibliography{biblio}

\begin{thebibliography}{10}
\providecommand{\url}[1]{{#1}}
\providecommand{\urlprefix}{URL }
\expandafter\ifx\csname urlstyle\endcsname\relax
  \providecommand{\doi}[1]{DOI \discretionary{}{}{}#1}\else
  \providecommand{\doi}{DOI \discretionary{}{}{}\begingroup
  \urlstyle{rm}\Url}\fi

\bibitem{LOI}
S.~Adrián-Martínez, et~al., J. Phys. G: Nucl. and Part. Phys. \textbf{43}(8),
  084001 (2016).
\newblock \doi{10.1088/0954-3899/43/8/084001}

\bibitem{muons}
M.~Ageron, et~al., Eur. Phys. J. C \textbf{80}(2), 99 (2020).
\newblock \doi{10.1140/epjc/s10052-020-7629-z}

\bibitem{runbyrun}
A.~Albert, et~al., J. C. A. P. \textbf{2021}(01), 064 (2021).
\newblock \doi{10.1088/1475-7516/2021/01/064}

\bibitem{angle}
R.~Abbasi, et~al., Phys. Rev. D \textbf{87}(1), 012005 (2013).
\newblock \doi{10.1103/PhysRevD.87.012005}

\bibitem{clark}
G.W. Clark, Phys. Rev. D \textbf{108}(2), 450 (1957).
\newblock \doi{10.1103/PhysRev.108.450}

\bibitem{IC_sun}
M.~Aartsen, et~al., Phys. Rev. D \textbf{103}(4), 042005 (2021).
\newblock \doi{10.1103/PhysRevD.103.042005}

\bibitem{antares_moon}
A.~Albert, et~al., The Eur. Phys. J. C \textbf{78}(12), 1006 (2018).
\newblock \doi{10.1140/epjc/s10052-018-6451-3}

\bibitem{antares_sun}
A.~Albert, et~al., Phys. Rev. D \textbf{102}(12), 122007 (2020).
\newblock \doi{10.1103/PhysRevD.102.122007}

\bibitem{ambrosio}
M.~Ambrosio, et~al., Astrop. Phys. \textbf{20}(2), 145 (2003).
\newblock \doi{10.1016/S0927-6505(03)00169-5}

\bibitem{achard}
P.~Achard, et~al., Astrop. Phys. \textbf{23}(4), 411 (2005).
\newblock \doi{10.1016/j.astropartphys.2005.02.002}

\bibitem{adamson}
P.~Adamson, et~al., Astrop. Phys. \textbf{34}(6), 457 (2011).
\newblock \doi{10.1016/j.astropartphys.2010.10.010}

\bibitem{bartoli}
B.~Bartoli, et~al., Phys. Rev. D \textbf{84}(2), 022003 (2011).
\newblock \doi{10.1103/PhysRevD.84.022003}

\bibitem{abeysekara}
A.U. Abeysekara, et~al., Proc. 33rd ICRC, Rio de Janeiro, Brazil  (2013).
\newblock \doi{10.48550/ARXIV.1310.0072}

\bibitem{dom}
S.~Aiello, et~al., J. Instr. \textbf{17}(07), P07038 (2022).
\newblock \doi{10.1088/1748-0221/17/07/P07038}

\bibitem{reco}
K.~Melis, A.~Heijboer, M.~De~Jong, et~al., Proc. 35th ICRC p. 950 (2017).
\newblock \doi{10.22323/1.301.0950}

\bibitem{astropy}
A.M. Price-Whelan, et~al., Astron. J. \textbf{156}(3), 123 (2018).
\newblock \doi{10.3847/1538-3881/aabc4f}

\bibitem{ICRS}
F.~Arias, P.~Charlot, M.~Feissel, J.~Lestrade, Astron. and Astrop.
  \textbf{303}, 604 (1995)

\bibitem{mupage}
G.~Carminati, M.~Bazzotti, A.~Margiotta, M.~Spurio, Comput. Phys. Com.
  \textbf{179}(12), 915 (2008).
\newblock \doi{10.1016/j.cpc.2008.07.014}

\bibitem{mupage2}
Y.~Becherini, A.~Margiotta, M.~Sioli, M.~Spurio, Astropart. P. \textbf{25}(1),
  1 (2006).
\newblock \doi{10.1016/j.astropartphys.2005.10.005}

\bibitem{pois}
S.~Baker, R.D. Cousins, Nucl. Instr. Meth. \textbf{221}(2), 437 (1984).
\newblock \doi{10.1016/0167-5087(84)90016-4}

\bibitem{sun_variation}
{Becker Tjus, J.}, {Desiati, P.}, {D\"opper, N.}, {Fichtner, H.}, {Kleimann,
  J.}, {Kroll, M.}, {Tenholt, F.}, A.\&A. \textbf{633}, A83 (2020).
\newblock \doi{10.1051/0004-6361/201936306}

\bibitem{cal}
R.~Coniglione, A.~Creusot, I.~Di~Palma, D.~Guderian, J.~Hofestaedt,
  G.~Riccobene, A.~Sánchez-Losa, Proc. 36th ICRC \textbf{ICRC2019}, 868
  (2019).
\newblock \doi{10.22323/1.358.0868}

\end{thebibliography}


\end{document}